\pdfoutput=1
\documentclass[a4paper,aps,american,citeautoscript,floatfix,pra,pdftex,superscriptaddress,%
longbibliography,%
reprint,twocolumn,final% add final to see real layout (e.g., no todos anymore)
]{revtex4-1}
\usepackage{amsfonts,amsmath,amssymb}
\usepackage[T1]{fontenc}
\usepackage[utf8]{inputenc}
\usepackage{graphicx}
\usepackage{microtype}
\usepackage[obeyFinal,textsize=footnotesize]{todonotes}
\usepackage{hyperref,hypernat}
\usepackage[todos]{CMImacros}

\newlength{\figwidth}
\newcommand*{\costau}{\ensuremath{\left<\cos^2\!\tau\right>}\xspace}
\newcommand*{\suppmat}{Supplemental Material~\cite{Note1}\xspace}
\newcommand*{\suppmatref}[1]{\jknote{check before submission}#1 in the \suppmat}

\makeatletter%
\renewcommand*{\@fnsymbol}[1]{\ensuremath{\ifcase#1\or \|\or *\or **\or \mathparagraph\or
      \mathsection\or \dagger\or \ddagger\or \dagger\dagger \or \ddagger\ddagger \else\@ctrerr\fi}}
\makeatother

\newcommand{\cfeldesy}{\affiliation{Center for Free-Electron Laser Science, Deutsches
      Elektronen-Synchrotron DESY, Notkestrasse 85, 22607 Hamburg, Germany}}%
\newcommand{\uhhcui}{\affiliation{The Hamburg Center for Ultrafast Imaging, Universität Hamburg,
      Luruper Chaussee 149, 22761 Hamburg, Germany}}%
\newcommand{\uhhphys}{\affiliation{Department of Physics, Universität Hamburg, Luruper Chaussee 149,
      22761 Hamburg, Germany}}%
\newcommand{\granada}{\affiliation{Instituto Carlos I de F\'{\i}sica Te\'orica y Computacional and
      Departamento de F\'{\i}sica At\'omica, Molecular y Nuclear, Universidad de Granada, 18071
      Granada, Spain}}%
\newcommand{\rgfemail}{\email[]{rogonzal@ugr.es}}%
\newcommand{\jkemail}{\email[]{jochen.kuepper@cfel.de}}%
\newcommand{\cmiweb}{\homepage{https://www.controlled-molecule-imaging.org}}%

\begin{document}
\title{Analyzing laser-induced alignment of weakly-bound molecular aggregates}
\author{Linda V. Thesing}\cfeldesy\uhhcui\uhhphys%
\author{Andrey Yachmenev}\cfeldesy\uhhcui%
\author{Rosario Gonz{\'a}lez-F{\'e}rez}\rgfemail\granada%
\author{Jochen Küpper}\jkemail\cmiweb\cfeldesy\uhhcui\uhhphys%
\date{\today}%
\begin{abstract}\noindent%
   The rotational and torsional dynamics of the prototypical floppy \indolew molecular cluster was
   theoretically and computationally analyzed. The time-dependent Schrödinger equation was solved
   for a reduced-dimensionality description of the cluster, taking into account overall rotations
   and the internal rotation of the water moiety. Based on our results, it became clear that
   coupling between the internal and the overall rotations are small, and that for typical field
   strengths in alignment and mixed-field orientation experiments the rigid rotor approximation can
   be employed to describe the investigated dynamics. Furthermore, the parameter space over which
   this is valid and its boundaries where the coupling of the internal and overall rotation can no
   longer be neglected were explored.
\end{abstract}
\maketitle

\section{Introduction}%
Biological function is strongly shaped by the intricate interaction of the molecules with their
aqueous environment. Unraveling the underlying (bio)molecule-water solvation interactions as well as
their relevance for chemical dynamics promises a detailed understanding of their contributions to
function. Approaching this through studies of the elementary chemical processes as intrinsic
properties in well-defined molecular aggregates enables the definition of fundamental building
blocks as a dynamical basis of intermolecular solute-solvent interactions and their chemical
dynamics. This rationalizes the longstanding history of detailed studies of molecule-solvent
clusters in the gas phase~\cite{Zwier:ARPC47:205, Cremer:NatChem10:8}.

Novel imaging techniques with highest spatiotemporal resolution, such as ultrafast
x-ray~\cite{Spence:PTRSB369:20130309, Kuepper:PRL112:083002} or electron
diffraction~\cite{Hensley:PRL109:133202, Yang:PRL117:153002}, photoelectron
imaging~\cite{Meckel:Science320:1478, Bisgaard:Science323:1464, Holmegaard:NatPhys6:428}, and
laser-induced electron diffraction~\cite{Blaga:Nature483:194} will provide a new level of detail to
these investigations and promise to allow for the recording of molecular movies of the dynamical
interactions. The applicability of these imaging methods to complex molecular systems relies on the
preparation of pure samples~\cite{Filsinger:PCCP13:2076, Chang:IRPC34:557} and benefits tremendously
from fixing the molecules in space~\cite{Spence:PRL92:198102, Filsinger:PCCP13:2076,
   Barty:ARPC64:415, Reid:PTRSA376:20170158}, \ie, to align or orient
them~\cite{Stapelfeldt:RMP75:543, Ghafur:NatPhys5:289, Holmegaard:PRL102:023001,
   Filsinger:JCP131:064309}. Recently, some of us have demonstrated the preparation of pure beams of
the prototypical \indolewater dimer cluster as well as its laser
alignment~\cite{Trippel:PRA86:033202, Trippel:arXiv:1802.04053, Trippel:JCP148:101103}.

However, it is not clear in how far the very floppy structure of weakly bound molecular clusters
modifies or hinders the control techniques, especially regarding alignment with strong laser fields.
It is well understood that internal rotation, or torsions, and overall rotation are
coupled~\cite{Gordy:MWMolSpec} and that internal rotations can also be controlled with the same
strong laser fields~\cite{Madsen:PRL102:073007}. So far, experimental and theoretical studies were
limited to highly symmetric molecular systems~\cite{Coudert:PRL107:113004, Madsen:PRL102:073007,
   Ramakrishna:PRL99:103001, Grohmann:PRL118:203201, Christensen:PRL113:073005}, \eg, with $G_{16}$
symmetry, such as biphenyls. It is \emph{a priori} not clear how these effects will transfer to
complex ``real world'' (bio)molecules and their complexes.

\begin{figure}
   \centering%
   \includegraphics[width=0.8\linewidth]{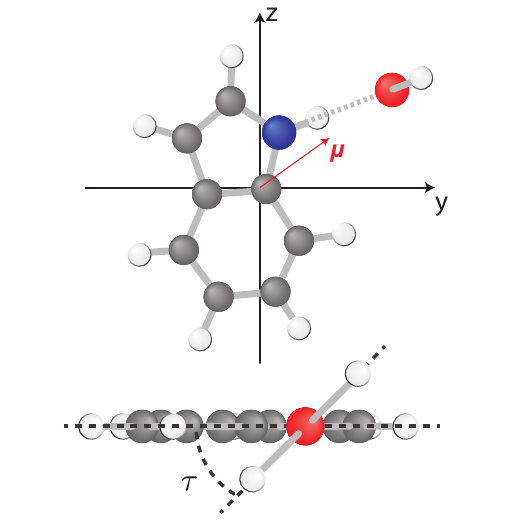}
   \caption{Sketch of the \indolew dimer cluster. The most polarizable axis defines the $z$-axis of
      the molecular frame. The torsional angle is defined as dihedral angle between the indole and
      water planes. For the experimentally determined structure see
      Ref.~\onlinecite{Korter:JPCA102:7211}.}
   \label{fig:indole-water}
\end{figure}

Here, we set out to analyze the laser alignment, and the corresponding influence of internal
rotations to the overall rotational dynamics, of molecule-solvent systems, which, generally, have
lower symmetries and asymmetric shapes of the constituents. Specifically, we start these
investigations with a theoretical analysis of the laser alignment of the prototypical \indolewater
dimer systems~\cite{Korter:JPCA102:7211, Carney:JCP108:3379, Mons:JPCA103:9958, Blanco:JCP119:880},
which we treat as a semi-rigid rotor with an additional one-dimensional internal rotation coordinate
corresponding to the rotation of the water moiety about its $b$-axis; see
\autoref{fig:indole-water}.
We utilize a reduced-mode variational approach based on the general-molecule variational
approach~\cite{Sorensen:LargeAmplitudes:1979, Matyus:JCP130:134112, Yurchenko:JMS245:126} combined
with a general treatment of electric fields~\cite{Owens:JCP148:124102}.

\section{Theory and computational setup}
The model structure of \indolew that is employed in this work is schematically shown in
\autoref{fig:indole-water}. The water molecule is attached to the planar indole frame \emph{via} a
hydrogen bond where the oxygen atom of the water molecule lies in the indole planer. We treat the
\indolew cluster as a floppy molecule with the water molecule undergoing an internal rotation. The
angle of internal rotation $\tau$ is defined as the dihedral angle between the indole and the water
planes, see \autoref{fig:indole-water}, with $\tau=\degree{90}$ in the equilibrium
configuration~\cite{Korter:JPCA102:7211}. We keep all other vibrational coordinates fixed at their
equilibrium values along the minimum energy path (MEP) of the internal rotation. The MEP is
determined by optimizing the structural parameters of \indolew complex at different values of the
$\tau$ coordinate between $\degree{0}$ and $\degree{360}$.

We assume the Born-Oppenheimer approximation and consider four degrees of freedom, three Euler
angles ($\phi$, $\theta$, $\chi$) describing the overall rotation of the system and the angle $\tau$
associated with the internal rotation of the water molecule. The field-free Hamiltonian of the
system is
\begin{eqnarray}
  H_0 &=& + \frac{1}{2}\sum_{\alpha,\beta=x,y,z} \hat{J}_\alpha G^{\text{rot}}_{\alpha\beta}(\tau) \hat{J}_\beta
   +\frac{1}{2}p_\tau G_\tau^{\text{tor}}(\tau) p_\tau \\ \nonumber
  &+& \frac{1}{2}\sum_{\alpha=x,y,z}
  \left[p_\tau G^{\text{cor}}_{\tau\alpha}(\tau) \hat{J}_\alpha
  + \hat{J}_\alpha G^{\text{cor}}_{\alpha\tau}(\tau) p_\tau
  \right]
  + V(\tau),
   \label{eqn:H_field_free}
\end{eqnarray}
where $\hat{J}_\alpha$ are components of the rotational angular momentum operator in the
molecule-fixed frame and $p_\tau=-i\hbar\partial/\partial\tau$. The kinetic energy matrices
$G_\tau^{\text{tor}}(\tau)$ and $G^{\text{rot}}_{\alpha\beta}(\tau)$ are associated with the
internal torsional and overall rotational motions, $G^{\text{cor}}_{\tau\alpha}(\tau)$ describes
coupling between the two motions, and $V$ is the potential energy function. The elements of the
kinetic energy matrices are calculated as functions of the torsional coordinate $\tau$ following the
generalized procedure from TROVE~\cite{Yurchenko:JMS245:126, Yachmenev:JCP143:014105}. The kinetic
energy matrices and potential energy function in \eqref{eqn:H_field_free} were built along the MEP.
The geometry optimizations were carried out using the density-fitted second-order M{\o}ller-Plesset
perturbation theory DF-MP2 in the frozen-core approximation, in conjunction with the augmented
correlation-consistent basis set aug-cc-pVTZ~\cite{Dunning:JCP90:1007, Kendall:JCP96:6796}. For the
density fitting approximation we utilized the JKFIT~\cite{Weigend:PCCP4:4285} and
MP2FIT~\cite{Haettig:PCCP7:59} auxiliary basis sets specifically matched to aug-cc-pVTZ. All
electronic structure calculations were carried out using Psi4~\cite{Parrish:JCTC13:3185}. The
\textit{ab initio} potential energy surface (PES) is represented by an analytical function by
fitting the expression
\begin{equation}
   V(\tau) = \sum_{n=0}^2 V_{2n} \cos(2n\tau)\ .
   \label{eqn:potential}
\end{equation}
The PES is depicted in \autoref[(a)]{fig:energy_levels} and the coefficients are
$V_0=-101.950~\invcm$, $V_2=106.006~\invcm$ and $V_4=-3.58143~\invcm$, in good agreement with the
experimental value for this motion of $V_2=99~\invcm$~\cite{Korter:JPCA102:7211}. The structural
parameters are represented by similar analytical functions; see the \suppmat for the \emph{ab
   initio} results of the EDM and polarizability, the analytical expressions and coefficients for
the internal coordinates, EDM and polarizability, as well as the symmetry properties of
indole(H$_2$O) in the presence of external electric fields~\footnote{See Supplemental Material
   for \emph{ab initio} results, analytical expressions, and symmetry properties.}.
\begin{figure}
   \includegraphics[width=\linewidth]{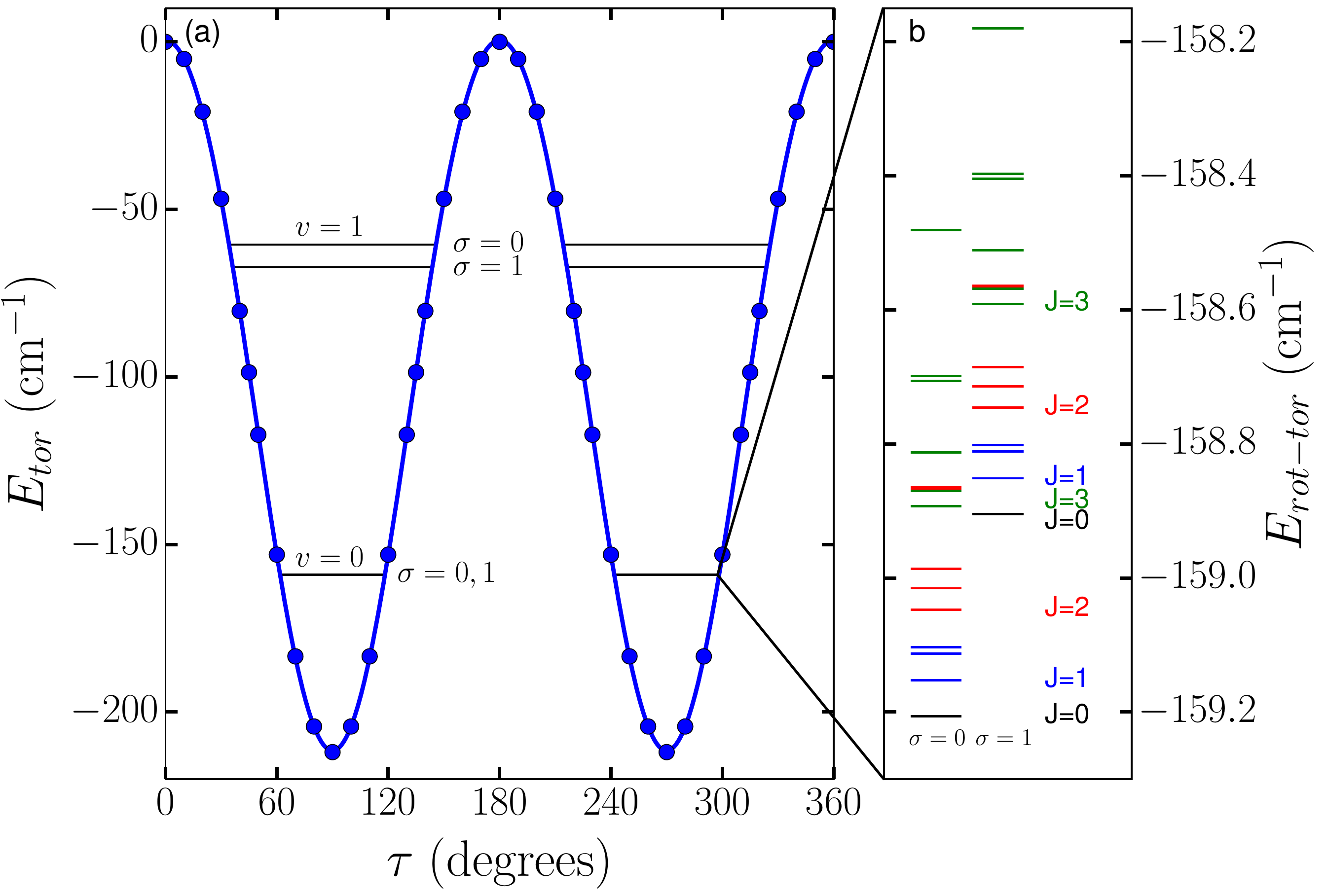}
   \caption{(a) The \textit{ab initio} 1D potential energy surface (blue circles) and corresponding
      fit (blue line) and the lowest two field-free torsional energy levels of \indolew, obtained
      from the pure torsional part of the Hamiltonian~\eqref{eqn:H_field_free}. Each torsional
      level, denoted by the vibrational quantum number $v$, is split into two sublevels $\sigma=0,1$
      of opposite parity. (b) Field-free rotation-torsional energy levels for $J=0,\ldots,3$
      corresponding to the torsional ground state, obtained from the
      Hamiltonian~\eqref{eqn:H_field_free}. Due to the small coupling between the internal and
      overall rotation, the energy levels are approximately given by the sum of the pure torsional
      and pure rotational energies, see text for more details.}
   \label{fig:energy_levels}
\end{figure}

Our analysis is restricted to a non-resonant linearly-polarized laser ac electric field combined
with a parallel weak dc electric field. The interaction of the polarizability with the weak dc field
is neglected. In addition, we can average over the rapid oscillations of the non-resonant ac field
and the interaction of the electric dipole moment with the laser field vanishes. The interaction of
the molecule with the external electric fields then reads
\begin{equation}
   H_\text{int} (t) = {-\boldsymbol{\mu}(\tau) \cdot  \Estat
      - \frac{1}{4} \Elaser(t)\underline{\underline{\alpha}}(\tau)\Elaser(t)}
   \ ,
   \label{eqn:H_interaction}
\end{equation}
where \Estat is the static electric field and $\Elaser(t)$ is the envelope of the laser electric
field. The electric dipole moment (EDM) $\boldsymbol{\mu}(\tau)$ and the polarizability tensor
$\underline{\underline{\alpha}}(\tau)$ of \indolew are calculated along the minimum energy path
created by varying the $\tau$ coordinate. They are represented by analytical functions similar to
the ones used for the PES, see \suppmat. The polarization axis of the laser is chosen as the
$Z$-axis of the laboratory fixed frame (LFF). The molecule fixed frame $(x,y,z)$ is defined by the
principle axes of polarizability at the equilibrium configuration so that the diagonal elements
fulfill $\alpha_{xx}<\alpha_{yy}<\alpha_{zz}$.

To study the rotational dynamics of \indolew, we solve the time-dependent Schrödinger equation
(TDSE) for the full Hamiltonian
\begin{equation}
   H (t) = H_0+H_\text{int} (t)\ ,
   \label{eqn:hamiltonian}
\end{equation}
using the short interative Lanczos method~\cite{Leforestier:JCOP94:59, Beck:PhysRep324:1} for the
time-propagation and a basis set expansion for the spatial coordinates using the eigenstates of the
field-free Hamiltonian~\eqref{eqn:H_field_free}. The field-free eigenbasis is calculated
variationally in several steps. First, the pure rotational and torsional basis functions,
$\Psi_l^{rot}(\phi, \theta, \chi)$ and $\Psi_l^{tor}(\tau)$, are constructed by diagonalizing the
respective parts of the field-free Hamiltonian~\eqref{eqn:H_field_free}. In this step, we use Wang
states~\cite{Wang:PR34:243}, \ie, symmetrized combinations of symmetric-top functions, as a basis
set for the rotational coordinates. For the torsional coordinates, we use sine and cosine functions
$\phi_n^\text{even}(\tau)=\cos(n\tau)/\sqrt{\pi}$ and
$\phi_n^\text{odd}(\tau)=\sin(n\tau)/\sqrt{\pi}$, $n>0$ and $\phi_0^\text{even}(\tau)=1/\sqrt{2\pi}$
to preserve the even and odd symmetry of states. The matrix elements of the full
Hamiltonian~\eqref{eqn:hamiltonian} are then set up in the product basis
$\Psi_l^{rot}(\phi, \theta, \chi)\Psi_m^{tor}(\tau)$ and transformed to the eigenbasis of the
complete field-free Hamiltonian~\eqref{eqn:H_field_free} following the generalized approach
developed in richmol~\cite{Owens:JCP148:124102}.

To analyze the importance of the internal motion of the water molecule, we compare our results to
calculations using the rigid rotor approximation~\cite{Omiste:PRA94:063408}, \ie, considering only
the three degrees of freedom of the overall rotation in~\eqref{eqn:H_field_free}. For the rigid
rotor calculations, we use the structural parameters, EDM and polarizability tensor at the
equilibrium configuration, \ie, $\tau=\degree{90}$.

\section{Results and Discussion}
In this section, we analyze the overall rotational and torsional dynamics of \indolew. We solve the
TDSE using the field-free ground state as the initial state of the time-propagation, \ie,
$T=0~\text{K}$. For the laser pulse, we consider a Gaussian envelope
$\Elaser(t)=\hat{\textbf{e}}_Z\textup{E}_0\exp\left(-4\ln2t^2/\tau_\text{FWHM}^2\right)$ with
$\textup{E}_0=2.74\times10^{7}~\Vpcm$ and $\tau_\text{FWHM}=1~\text{ns}$. The parallel static
electric field is increased to a field strength of $\Estat=600~\Vpcm$ slowly enough to ensure
adiabatic behavior before the laser field is applied. We analyze the alignment and mixed-field
orientation dynamics of \indolew, quantified by the expectation values \costhreeD and \costheta,
respectively, as well as the torsional alignment \costau.

\begin{figure}
   \includegraphics[width=\linewidth]{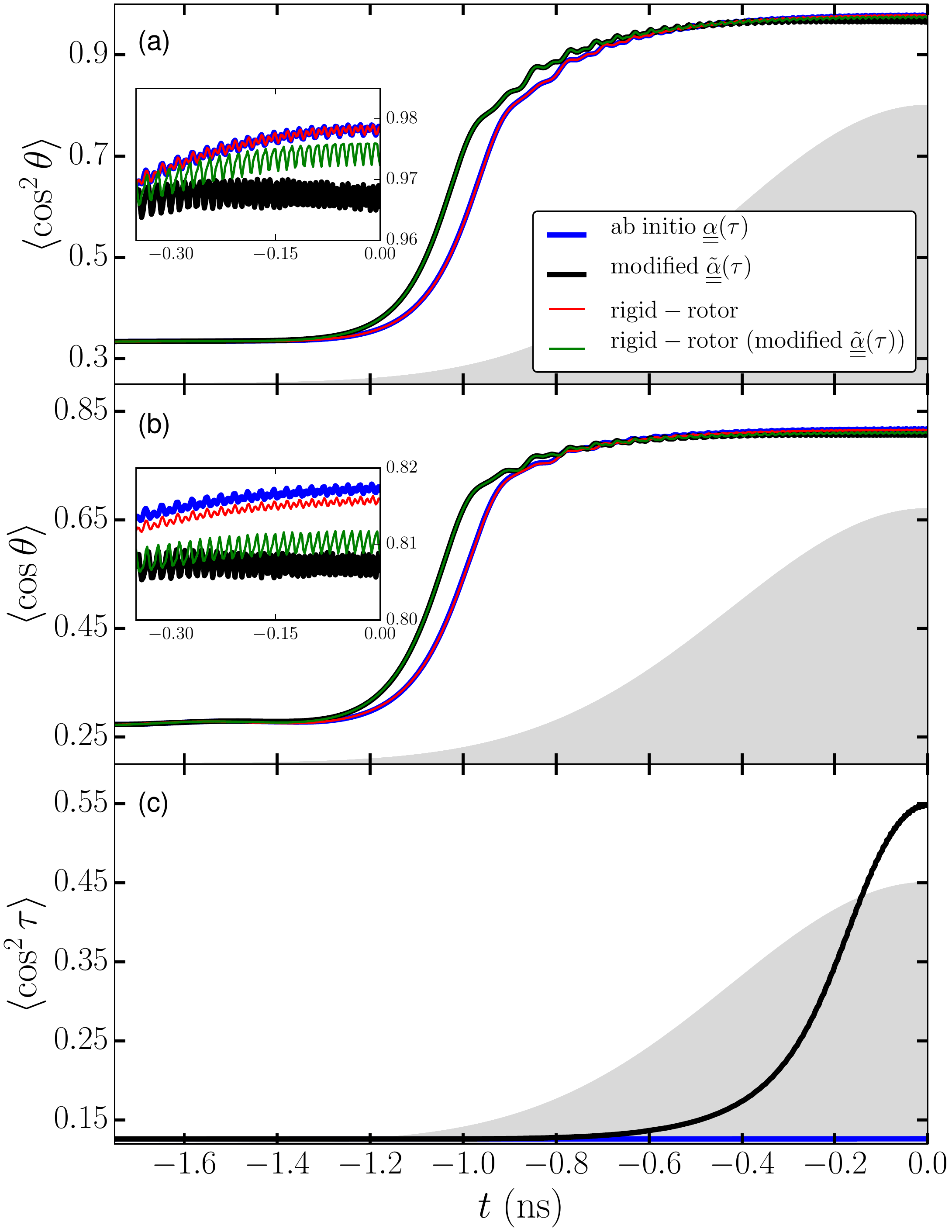}
   \caption{For the ground state of \indolew, the time evolution of the expectation values (a)
      \costhreeD, (b) \costheta and (c) \costau using $\underline{\underline{\alpha}}(\tau)$ from
      \textit{ab initio} calculations (thick blue lines), the modified
      $\tilde{\underline{\underline{\alpha}}}(\tau)$ (thick black lines) and using the rigid rotor
      approximation (thin red lines). The results for the \textit{ab initio} polarizability
      $\underline{\underline{\alpha}}(\tau)$ and the rigid rotor approximation in (a) and (b) are
      practically indistinguishable. The result with the smallest value of \costhreeD and \costheta
      and the largest value of \costau at the peak intensity are obtained for the modified
      polarizability $\tilde{\underline{\underline{\alpha}}}(\tau)$. The insets in (a) and (b) show
      a zoom of \costhreeD and \costheta close to the peak intensity. The grey area illustrates the
      envelope of the laser field with $\textup{E}_0 = 2.74\times10^{7}~\Vpcm$. The static field
      strength is $\Estat=600~\Vpcm$.}
   \label{fig:results}
\end{figure}
\autoref[(a) and (b)]{fig:results} show the time evolution of the alignment and orientation of the
ground state of \indolew obtained using the non-rigid rotational-torsional Hamiltonian (blue lines)
as well as the results using the rigid rotor approximation (red lines). As the laser field strength
rises, the most polarizable axis (MPA) becomes strongly aligned along the polarization axis of the
laser, reaching $\costhreeD=0.98$ at the peak intensity. For both, the rigid and the non-rigid
models, we observe a typical mixed-field-orientation dynamics that has previously been described for
other linear and asymmetric top molecules~\cite{Nielsen:PRL108:193001, Trippel:PRL114:103003,
   Omiste:PRA86:043437, Omiste:PRA88:033416}. Due to the presence of the static electric field, the
orientation increases as laser-aligned pendular states are formed and increasingly coupled with
increasing laser field strength. The orientation at the peak intensity, $\costheta=0.82$, is lower
than the value obtained in the adiabatic description, $\costheta=0.99$, indicating a non-adiabatic
dynamics, \ie, several field-dressed eigenstates of the instantaneous
Hamiltonian~\eqref{eqn:hamiltonian} contribute to the time-dependent wave
function~\cite{Omiste:PCCP13:18815, Omiste:PRA94:063408, Thesing:JCP146:244304}.

We point out that the results obtained for the rigid and non-rigid descriptions are practically
indentical, which shows that the \indolew cluster can be treated as a rigid molecule for moderate
electric field strengths that are typically employed in molecular alignment and mixed-field
orientation experiments. To understand why the internal rotation of the water molecule does not
influence the overall rotational dynamics, we look at the coupling between the internal and overall
rotations and the resulting rotation-torsional energy levels. In the field-free case, the coupling
of the two motions is described by the Coriolis-type coupling and the dependence of the kinetic
energy matrix $G^{\text{rot}}_{\alpha\beta}(\tau)$ on the torsional angle $\tau$ in
\eqref{eqn:H_field_free}. Since this field-free coupling is very small~\cite{Korter:JPCA102:7211},
the eigenstates of the Hamiltonian~\eqref{eqn:H_field_free} are approximately described by the
product states $\Psi_l^{rot}(\phi,\theta,\chi)\Psi_m^{tor}(\tau)$, and the energy levels are given
approximately by the sum of the pure rotational and torsional energies, \autoref{fig:energy_levels},
even for large values of the quantum number $J$. Due to tunneling, each torsional level splits into
two sublevels, denoted by $\sigma=0,1$ in \autoref[(a)]{fig:energy_levels}. The energy difference
between two consecutive torsional states is three orders of magnitude larger than the energy gaps
between pure rotational states. As a result, the rotation-torsion energy levels are distributed as
bands of rotational states for each torsional sublevel, see~\autoref[(b)]{fig:energy_levels}.

In the presence of an external electric fields, the internal and overall rotations are additionally
coupled due to the dependence of the EDM and the polarizability on the torsional angle, see
\eqref{eqn:H_interaction}, which is weak for \indolew. Thus, for the ac and dc field strengths
considered here, the field-induced coupling between the torsional ground and first excited state
$\braopket{\Psi^{rot-tor}_{i,v=0}}{H_\text{int}}{\Psi^{rot-tor}_{j,v=1}}$ where $v=0,1$ indicates
the torsional ground and excited state, respectively, is small ($<1~\invcm$) compared to the energy
gap $\Delta{E}\approx98~\invcm$ between these torsional levels. Due to the symmetry of the EDM and
the polarizability, the two sublevels $\sigma=0,1$ of a torsional state are not coupled by the
external electric fields, see \suppmatref{Section~IV}. As a consequence, the field-dressed
rotational-torsional wavepacket is dominated by the torsional ground state. The torsional alignment
shown in \autoref[(c)]{fig:results} (blue lines) thus remains constant with $\costau=0.126$. To
achieve a field-induced coupling that is strong enough to overcome the energy gap between the
torsional ground and first excited state, laser field strengths larger than
$\textup{E}_0\approx10^{8}~\Vpcm$, \ie, intensities $\Icontrol>10^{13}~\Wpcmcm$, would be necessary.
Such fields would, however, affect the electronic structure and induce ionization of \indolew and
are generally not used to control the rotational dynamics of these
clusters~\cite{Trippel:JCP148:101103}.

In the following, we investigate for which regimes of field-free and field-induced couplings the
rigid rotor approximation can still be applied to describe the rotational dynamics. To obtain a
large field-induced coupling, stronger external ac and dc electric fields could be applied or the
dependence of the EDM and the polarizability on the torsional angle $\tau$ could be, artificially,
increased. We begin by studying a molecular cluster that has a modified polarizability
$\tilde{\underline{\underline{\alpha}}}(\tau)$ and is otherwise identical to \indolew. Since the
interaction with the weak static electric field is much weaker than the one with the laser field, we
do not modify the EDM. We set $\tilde{\alpha}_{pq}(\tau)$ including only terms with $n\leq1$ and
with the coefficients listed in \suppmatref{Table~V}. The coefficients $\tilde{\alpha}_2^{(pq)}$ are
chosen to be $200$ times larger than the largest ones obtained from fits to the ab inito results.
The chosen $\tilde{\alpha}_0^{(pq)}$ satisfy
$\tilde{\underline{\underline{\alpha}}}(\tau=\degree{90})=\underline{\underline{\alpha}}(\tau=\degree{90})$.
This polarizability increases the field-induced coupling by a similar magnitude as increasing the
laser field strength by a factor of $\sqrt{200}$. We carry out an additional rigid rotor calculation
using the structural parameters and EDM at the equilibrium configuration, but the expectation values
$\expectation{\tilde{\alpha}_{pq}}=\braopket{\Psi_0^{tor}}{\tilde{\alpha}_{pq}(\tau)}{\Psi_0^{tor}}$
in the torsional ground state as these values differ significantly from
$\tilde{\alpha}_{pq}(\tau=\degree{90})$. We point out that the expectation value of the modified
polarizability is not diagonal in the chosen MFF, \ie, the MPA of the modified polarizability is not
parallel to the MFF $z$-axis.

The results for this enhanced-response molecular system are depicted in \autoref{fig:results} for
the non-rigid (black lines) and rigid rotor cases (green lines). At lower laser field strengths, the
alignment does not differ from the results obtained using the rigid-rotor approximation. Close to
the peak intensity, the alignment starts to decrease and reaches a value of \costhreeD=0.965,
slightly smaller than the rigid rotor result \costhreeD=0.972. Simultaneously, the orientation
starts to differ slightly from the rigid rotor result reaching a smaller value at the peak
intensity. Regarding the torsional alignment, we observe an increase of \costau with increasing
laser field strength, due to the contribution of excited torsional states, in particular the second
excited state $v=2,\sigma=0$. The reason for this is that due to the symmetry of the polarizability,
the coupling between the field-free rotation-torsional states with different torsional symmetry is
small~\cite{Berden:JCP104:972, Trippel:PRA86:033202}, see the \suppmat. The difference between the
non-rigid and the rigid rotor description can be understood in terms of the effective polarizability
$\braopket{\Psi_m^{tor}}{\tilde{\underline{\underline{\alpha}}}(\tau)}{\Psi_m^{tor}}$ of the
torsional states. For weak laser fields, excited torsional states are not involved in the dynamics.
As a consequence, the alignment and orientation evolve in a similar way as for the rigid \indolew.
The torsional ground state wave function is localized around $\tau=\degree{90}$ yielding a
polarizability with small off-diagonal elements
$\expectation{\tilde{\alpha}_{13}}=\expectation{\tilde{\alpha}_{31}}=5.2~\au$;
$\tilde{\alpha}_{13}(\tau=\degree{90})=0$. At stronger laser fields, the contributions of excited
torsional states modify the polarizability, with $\expectation{\tilde{\alpha}_{13}}=23.5~\au$ at the
peak intensity. As a result, a different molecular axis is aligned compared to the rigid rotor case,
and \costhreeD decreases. In addition to the change in the MPA, the anisotropy of the polarizability
increases with increasing laser field strength. The alignment of the MPA
$\expectation{\cos^2\theta_\text{MPA}}$, where $\theta_\text{MPA}$ is the angle between the MPA and
the LFF $Z$-axis, is thus larger at the peak intensity than for the rigid rotor case. As a result of
the change in the MPA, the orientation \costheta in~\autoref[(b)]{fig:results} also decreases. Thus,
a larger orientation of the MPA $\expectation{\cos\theta_\text{MPA}}$ is obtained for the non-rigid
than for the rigid case.

Additionally modifying the EDM in an analogous way results in different effective dipole moments
$\expectation{\boldsymbol{\mu}}$ and a correspondingly changed degree of orientation compared to the
rigid-rotor result. However, since the interaction of the dc electric field with the dipole moment
is comparably weak, no additional excitation of torsional states occurs. Increasing the laser field
strength by the corresponding factor of $\sqrt{200}$ does not have the same impact on the rotational
dynamics as the modified polarizability. While a stronger laser field leads to contributions of
excited torsional states, the expectation value of the polarizability in these excited states does
not differ much from the one in the torsional ground state. Thus, no significant change of the MPA
or polarizability anisotropy occurs. Calculations performed for a laser field strength of
$\textup{E}_0=3\times10^{8}~\Vpcm$ and $\Estat=0$ show a small increase in \costau during the laser
pulse, but no difference in the overall alignment \costhreeD between the rigid and non-rigid
descriptions. Increasing the laser field strength further was computationally too expensive.

In addition to analyzing the impact of a strong field-induced coupling, we investigate the influence
of the barrier height. This affects the torsional energy level structure~\cite{Gordy:MWMolSpec} and
may thus alter the field-dressed coupling necessary to achieve an excitation of the torsion in the
presence of external fields. To this end, we use a modified torsional potential
$\tilde{V}(\tau)=\tilde{V}_0(\cos(2\tau)-1)$ with a very small barrier height
$\tilde{V}_0=1.0~\invcm$. Using the same field parameters as in~\autoref{fig:results} and the
\textit{ab initio} polarizability, we observe an overall rotational dynamics that is very similar to
the dynamics obtained using the \textit{ab initio} torsional potential. The reason for this is that
by lowering the barrier height the energy gap between the torsional ground and second excited level
is only decreased to $\Delta{E}\approx55~\invcm$, which is still large compared to the coupling
induced by the external electric fields. Finally, we consider molecules with an internal rotor that
has a smaller rotational constant. For a modified molecular system increasing the mass of the
protons in the water molecule of \indolew to $14~\text{u}$ and keeping the \textit{ab initio}
polarizability, the rotational dynamics for the rigid and non-rigid descriptions are almost
indistinguishable when applying the field configuration as in~\autoref{fig:results}. If we
additionally consider a low torsional barrier with $\tilde{V}_0=1.0~\invcm$ (see above), the energy
gap between the torsional ground state and second excited state is lowered to
$\Delta E \approx 4.5~\invcm$ and we observe contributions of excited torsional states. Here, the
field-free rotation-torsional states belonging to the excited torsional levels cannot be
approximated well as product states of pure rotational and torsional states. As a consequence, the
orientation slightly differs from the rigid rotor result with $\Delta\costheta=0.01$.

\section{Conclusion}
We investigate the rotational dynamics of a floppy molecule in combined laser and static electric
fields. Our work is focused on the prototypical \indolew cluster where the attached water molecule
undergoes an internal rotation. The molecular structure, electric dipole moment and polarizability
are calculated with \textit{ab initio} methods. We solve the time-dependent Schrödinger equation for
a moderate laser field strength and a weak dc electric field taking into account four degrees of
freedom for the internal and overall rotation. We compare the obtained alignment and orientation
dynamics to results computed within the rigid rotor approximation. We demonstrate that \indolew can
be treated as a rigid molecule in typical alignment and mixed-field orientation experiments. This
conclusion is rationalized by the weak field-free and field-induced couplings of the internal and
overall motions compared to the respective energy spacings. We explore regimes of laser-field
strengths for which the internal rotation of the water moiety can no longer be neglected. However,
such strong laser pulses, $\Icontrol>10^{13}~\Wpcmcm$, are not likely to be used in alignment
experiments as they would result in electronic excitation and ionization. Let us remark that even at
a rotational temperature of $T=0~\text{K}$, both torsional sublevels $\sigma=0,1$ of the rotational
ground state would be populated in an molecular beam according to their nuclear spin statistical
weights. However, as the rotational and torsional dynamics of these two sublevels in the presence of
the ac and dc fields do not differ significantly, our analysis of the $\sigma=0$ sublevel provides a
good description of the overall rotational dynamics in the external fields.

We analyze the influence of a larger field-induced coupling on the overall and internal rotational
dynamics by using artificially modified electric dipole moment and polarizability with a two orders
of magnitude stronger dependence on the torsional angle. We find a decrease of the overall alignment
and orientation close to the peak intensity due to an effective change in the MPA in the excited
torsional state. In addition, we observe an increase of the torsional alignment with increasing
laser field strength caused by the contributions of excited torsional states. In contrast, for
unmodified \indolew, a constant torsional alignment is found.

Based on these results for \indolew, we conclude that similar molecular clusters can also be treated
as rigid molecules. This conclusion can also be extended to other floppy molecular systems provided
their dipole moment and polarizability depends only weakly on the internal motion and their
rotational and torsional energy levels are approximately given by the sum of pure rotational and
torsional energies with large gaps between consecutive torsional levels. For molecules with a small
torsional barrier or a smaller internal rotational constant, \eg, for previously studied
biphenyl-type molecules, a weaker field induced coupling may be necessary to achieve contributions
of excited torsional states to the field-dressed dynamics. However, the rigid rotor approximation
can still be employed for moderate laser fields strengths. This is different than for previously
studied biphenyl-type molecules, with their axis of internal rotation parallel to a principle axis
of intertia and their MPA not modified by the torsion, which results in a qualitatively different
coupled rotational and torsional dynamics than for generic low-symmetry molecular clusters, such as
\indolew. To achieve an accurate description of the rotational dynamics for molecules with small
energy gaps between the torsional ground and excited states, the coupling of the internal and
overall rotations cannot be neglected. Furthermore, for impulsive alignment even the small coupling
between the internal and overall rotation in \indolew becomes important on longer, \eg, nanosecond,
timescales.

The variational approach applied in this work allows to extend our study to include multiple
internal modes in a multi-dimensional PES and to study their effect on the field-dressed rotational
dynamics, which will be investigated in the future.

\section{Acknowledgements}
This work has been supported by the Deutsche Forschungsgemeinschaft (DFG) through the excellence
cluster ``The Hamburg Center for Ultrafast Imaging -- Structure, Dynamics and Control of Matter at
the Atomic Scale'' (CUI, EXC1074) and through the priority program ``Quantum Dynamics in Tailored
Intense Fields'' (QUTIF, SPP1840, KU 1527/3), by the European Research Council under the European
Union's Seventh Framework Programme (FP7/2007-2013) through the Consolidator Grant COMOTION
(ERC-614507-Küpper), and by the Helmholtz Association ``Initiative and Networking Fund''. R.G.F.\
gratefully acknowledges financial support by the Spanish project FIS2014-54497-P (MINECO) and by the
Andalusian research group FQM-207 and the grant P11-FQM-7276.

\bibliography{string,cmi}
\onecolumngrid
\end{document}